\documentclass[prl,epsfig,twocolumn,amssymb]{revtex4}

\usepackage{graphicx}

\begin{document}
\title{Dynamics of a quantum phase transition in a ferromagnetic   Bose-Einstein 
condensate}
\begin{abstract}
We discuss dynamics of a slow quantum phase transition in a spin-1
Bose-Einstein condensate. We determine analytically the scaling properties of the
system magnetization and verify them with  numerical simulations in  a one
dimensional model. 
\end{abstract}
\author{Bogdan Damski and Wojciech H. Zurek}
\affiliation{
Theory Division, Los Alamos National Laboratory, MS-B213, Los Alamos, NM 87545, USA
}
\maketitle

Studies of phase transitions have  traditionally focused 
on {\it equilibrium} scalings of various properties near the critical point.
Dynamics of the phase transition presents new challenges and there is a strong motivation
for analyzing it. Nonequilibrium  phase 
transitions may play a role in  the evolution of the early Universe \cite{kibble}.
Their analogues can be studied in the condensed matter
experiments. The latter observation led to 
development of the theory based on the universality of  critical
behavior \cite{zurek}, which in turn resulted 
in a series of beautiful experiments \cite{eksperymenty}.
The recent progress in the cold atom experiments allows for 
time dependent realizations of different models undergoing a quantum phase
transition (QPT) \cite{lewenstein,nature_kurn}. 
These experimental developments are only a proverbial tip of the iceberg, but they call 
for an in-depth theoretical understanding of the QPT dynamics.

A QPT is a fundamental change in {\it ground state} (GS) of the system  as a result of
 small variations of an external parameter, e.g., 
 a magnetic field \cite{sachdev}.
It takes place  ideally at zero absolute temperature, which is in striking 
contrast to thermodynamical phase transitions. 
The most complete description of the QPT dynamics has been obtained
so far in  spin models
\cite{dorner,spiny} that are exactly solvable.
In these systems the gap in the
excitation spectrum goes to zero at the critical point, which precludes the 
adiabatic evolution across the phase boundary. It leads to creation of 
excitations whose density and scaling with a quench rate 
follow from a quantum version \cite{dorner,bodzio} of the Kibble-Zurek (KZ) theory
\cite{kibble,zurek}.

We study dynamics of a ferromagnetic condensate of 
spin-1 particles \cite{fer}. 
For simplicity, we consider 
1D homogeneous (untrapped) clouds: atoms in a box as in the 
experiment \cite{raizen} with spinless bosons.
We adopt  the parameters for our 1D  model 
such that the length and time scales are comparable to experimental 
ones \cite{units}. Assuming that the system is placed in a 
magnetic field $B$ aligned in the $z$ direction, one gets the following 
dimensionless mean-field energy functional \cite{units}
\begin{eqnarray}
\label{E}
E[\Psi]= \int&dz&\frac{1}{2}\frac{d\Psi^\dag}{dz}\frac{d\Psi}{dz}
       + \frac{c_0}{2}\left(\Psi^\dag\Psi\right)^2 
       + Q\langle\Psi|F_z^2|\Psi\rangle\nonumber\\&+&  
       \frac{c_1}{2}\sum_\alpha\langle\Psi|F_\alpha|\Psi\rangle^2
\end{eqnarray}
where 
$\Psi^T=(\psi_1,\psi_0,\psi_{-1})$ describes the $m=0,\pm1$ condensate components, 
$\int dz \sum_m|\psi_m|^2=1$, and
$F_{x,y,z}$ are spin-1 matrices \cite{ueda_broken}. 
The first term in (\ref{E}) is the kinetic energy,
the second and the fourth term describe spin-independent 
and spin-dependent atom interactions respectively, the third term is a
quadratic Zeeman shift coming from atom interactions with a magnetic field.
For $^{87}$Rb atoms considered
here $c_1<0$, which results in an interesting phase diagram due to 
the competition between the last two terms in (\ref{E}).
Restricting analysis  to zero longitudinal  magnetization case, 
and introducing  
$$
q=Q/(n|c_1|),\ \ n=\Psi^\dag\Psi
$$
one finds  a polar  phase for $q>2$, described by $\Psi_P^T\sim(0,1,0)$, 
and the broken-symmetry  phase where  
$$
\Psi_B^T\sim(\sqrt{4-2q}e^{i\chi_1},
\sqrt{8+4q}e^{i(\chi_1+\chi_{-1})/2},\sqrt{4-2q}e^{i\chi_{-1}})
$$
for $0\le q<2$.
The freedom of choosing the $\chi_{\pm1}$ results in rotational symmetry 
of the  transverse magnetization on the $(x,y)$ plane.
The transition between these phases can be
driven by the change of  the magnetic field $B$ imposed on the 
atom cloud, $q\sim Q\sim B^2$ \cite{ueda_pra2007}, which was experimentally 
done in \cite{nature_kurn}.

The dynamics of a QPT depends  on the rate of quench 
driving the system across the phase boundary. 
For very fast ``impulse'' transition, 
the system has no time to adjust to the changes of the Hamiltonian  and 
arrives in a region where a new phase is expected with 
the ``old'' wave-function untouched during the evolution.
Slow transitions are  different: the system has  time to ``probe'' 
various broken symmetry ``vacua'' in the neighborhood of the critical 
point where it gets excited.
We are interested in  evolutions that are fast enough
to produce macroscopic excitations of the system, but slow enough to 
reflect scalings of the critical region. By comparing analytical 
findings to numerical simulations for experimentally relevant parameters
we provide the first complete description of QPT dynamics in a ferromagnetic 
condensate.

Fast  transitions were realized in the Berkeley experiment
\cite{nature_kurn}. The 3D numerical simulations closely following this 
experiment were reported in \cite{ueda_pra2007}. Analytical studies 
of the evolution after ``impulse'' quench  were  
presented in \cite{austen,uwe}. The paper of  Lamacraft 
\cite{austen} also discusses dynamics of non instantaneous transitions 
in  2D spinor condensates focusing on analytical predictions on 
the growth of the transverse magnetization correlation functions. 

We start  with a qualitative discussion.
Considering  small perturbations around the GS of the broken-symmetry phase
one  finds  three Bogolubov modes as in \cite{ueda_broken}
where quantum fluctuations are studied.
In the long wavelength limit (important for slow transitions) 
there is only one nonzero eigenvalue mode:
the gapped mode having eigenenergy $\Delta\sim\sqrt{4-q^2}$. 
Suppose now that we drive the system from polar to broken-symmetry phase.
The system reaction time
to Hamiltonian changes in the broken-symmetry phase is given by the inverse 
of the gap: $\tau\sim\frac{1}{\Delta}$ \cite{dorner,bodzio}. For example, 
when $\tau$ is
small enough the evolution becomes adiabatic  so the system adjusts
fast  to parameter changes. Right after entering the broken-symmetry
phase, the reaction time is 
large with respect to the transition time,  $\Delta/\frac{d\Delta}{dt}$, and so
the system undergoes the ``impulse'' evolution where its state is ``frozen''.
The gapped mode starts to be populated around the instant $\hat t$ after
entering the broken-symmetry phase: the system leaves the ``impulse'' regime to 
catch up with  instantaneous GS solution. This happens when the two time scales 
become comparable: $1/\Delta(\hat t)\sim\Delta/\frac{d\Delta}{dt}|_{t=\hat t}$.
We consider here transitions driven by 
\begin{equation}
\label{q_od_t}
q(t)=2-t/\tau_Q,
\end{equation}
where $\tau_Q$ is the quench time inversely proportional to the speed of
driving the system through the phase transition. 
For slow transitions of interest here, $\tau_Q\gg1$, we obtain
\begin{equation}
\label{t_hat}
\hat t\sim \tau_Q^{1/3}.
\end{equation}

In the following we analyze  dynamics induced by a linear decrease of $q(t)$ (\ref{q_od_t}).
The evolution starts from $t<0$, i.e., in the polar phase, and ends 
at $t=2\tau_Q$ ($q=0$). Such $q(t)$ dependence  is achieved 
by ramping down the magnetic field as $\sim\sqrt{2-t/\tau_Q}$.
The initial state is chosen as a slightly 
perturbed GS in the polar phase,
$
\Psi^T\sim(\delta\psi_1, 1/\sqrt{L}+\delta\psi_0, \delta\psi_{-1}),
$
where $|\delta\psi_m|\ll 1/\sqrt{L}$ are random.  
We generate the real and imaginary
part of $\delta\Psi_m$ at  different grid points with the 
probability distribution $p(x)=\exp(-x^2/2\sigma^2)/\sqrt{2\pi}\sigma$. 
We take  $\sigma= 10^{-4}$  to start evolution
closely to the polar phase  GS.

\begin{figure}[t]
\includegraphics[width=\columnwidth,clip=true]{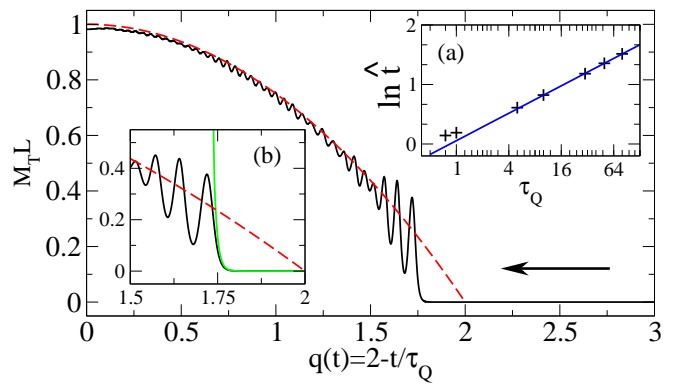}
\caption{(color) Main plot: numerical solution (black solid line) vs. 
static prediction (red dashed line). 
The arrow depicts direction of evolution.
Inset (a): the same 
as in the main plot plus a numerically obtained 
solution of the linearized problem (green divergent line). Inset (b): 
numerical data vs.  fit to $\tau_Q\ge10$ data only (see text for details). 
In the main plot $\tau_Q= 10$ (see \cite{units} for
units).
}
\label{fig1}
\end{figure}

To find the full  numerical solution within the mean-field approximation,
we integrate three coupled nonlinear Schr\"odinger 
equations for the $\psi_m$ condensates that can be easily obtained by the
variation of (\ref{E}). 
During evolution we look at the magnetization of the
sample 
$$
f_\alpha=\langle\Psi|F_\alpha|\Psi\rangle, \ \ \alpha=x,y,z.$$
{\it The transverse magnetization.} A total transverse (to the magnetic field
in the $z$ direction) magnetization reads
\begin{equation}
\label{MT}
M_T(t)=\int dz [f_x^2(z,t)+f_y^2(z,t)]=\int dz\,m_T,
\end{equation}
and is experimentally measurable. 
It disappears in the GS of the polar phase and 
equals $(1-q^2/4)/L$ in the broken-symmetry  GS.
Its  typical
evolution is depicted in Fig. \ref{fig1}. We see there that nothing happens 
in the polar phase. The system starts nontrivial evolution in the
broken-symmetry phase at a distance 
$\hat t/\tau_Q$ after the critical point was passed.
The magnetization grows  fast 
from that point until it exceeds the static
prediction and starts oscillations  with the amplitude decreasing 
in time. We consider slow transitions. Therefore, 
by the end of time evolution, when $q=0$, the system is in the slightly
perturbed  ferromagnetic GS: globally $M_TL\approx1$ (Fig. \ref{fig1}) 
and locally   $L^2m_T(z)\approx1$ (Fig. \ref{fig3}).
We can now ask:  Does the scaling (\ref{t_hat})  hold? 
To find out we define arbitrarily $\hat t$ as the instant when $M_TL$ intersects $1\%$.
A  fit to numerics for $\tau_Q\ge10$ yields   
$\ln\hat t=(0.056\pm0.01)+(0.332\pm0.002)\ln\tau_Q$ which confirms prediction 
(\ref{t_hat}). This fit is presented in Fig. \ref{fig1}a, where the gradual departure of the 
numerical data for $\tau_Q<10$ from 
$\hat t\sim\tau_Q^{1/3}$   indicates that
$\tau_Q\gg1$ or $37ms$ has to be taken for the observation of $1/3$ 
exponent: quench has to be slow enough to reflect the critical
dynamics. 
\begin{figure}[t]
\includegraphics[width=\columnwidth,clip=true]{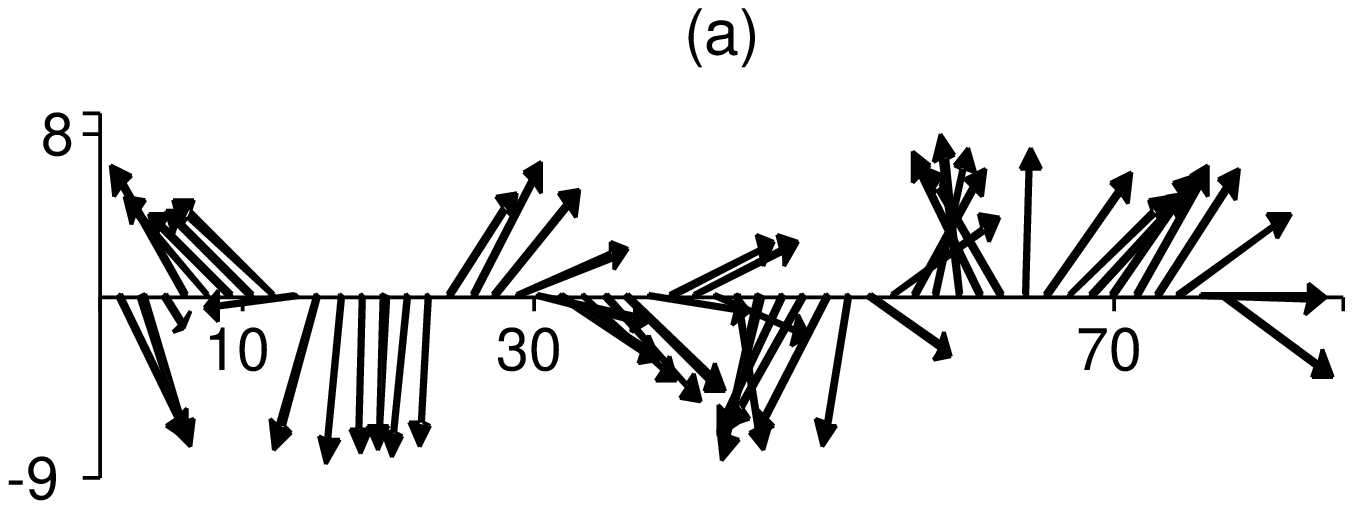}\\
\includegraphics[width=\columnwidth,clip=true]{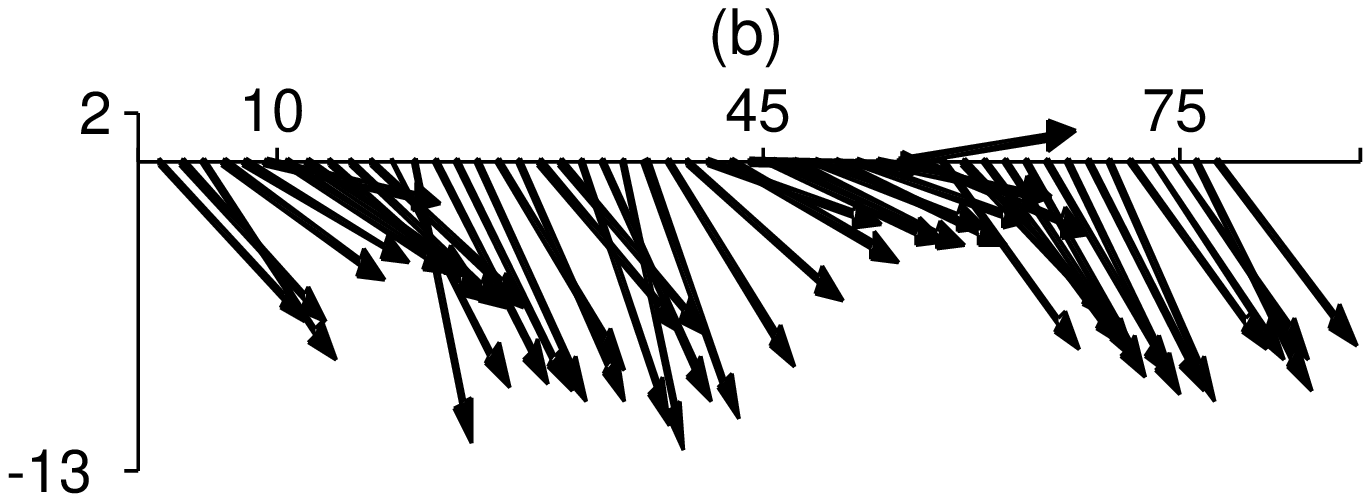}
\caption{The vectors represent $(f_x(z),f_y(z))\times10^3$. Plot (a): 
snapshot at $q(t=2.81)= 1.72$, i.e., at the first peak in 
$M_TL$ (see Fig. \ref{fig1}). Plot (b): snapshot by the 
end of time evolution: $q(t=20)=0$. The results come from the same numerical simulation
as in Fig. \ref{fig1} (see \cite{units}  for units). 
}
\label{fig2}
\end{figure}

In the GS configuration of the
broken-symmetry phase the vector $(f_x,f_y)$ can have arbitrary orientation,
so in the dynamical problem considered here it is interesting to find
out how is this  rotational symmetry  broken. 
When unstable evolution starts,  spatial correlations
in magnetization appear (Fig. \ref{fig2}a).
In the subsequent evolution these correlations  evolve such that the
correlation length increases: see Fig. \ref{fig2}b  obtained by the end of
time evolution. This is a generic picture though the details 
depend on the quench time  $\tau_Q$ and initial state of the system. 
This behavior suggests creation of spin textures \cite{stephens,vilenkin}. 
In our case, 
topological textures are  spin  configurations where
the magnetization direction varies in space so that the kinetic energy term in
(\ref{E}) is not minimized, but magnetization magnitude  
follows closely  a GS result.
Such structures appear in 1D when the first homotopy group of the
vacuum manifold $\cal M$ is nontrivial, which happens here:
$\pi_1({\cal M})={\mathbb Z}$ \cite{makela}. These textures are characterized 
by the winding number,
$
\frac{1}{2\pi}\int dz \frac{d}{dz}{\rm Arg}(f_x+if_y),
$
which is not conserved. Indeed, 
it reads $+1$ in Fig. \ref{fig2}a, while 
by the end of that evolution (Fig. \ref{fig2}b) it equals $0$.

\begin{figure}[t]
\includegraphics[width=\columnwidth,clip=true]{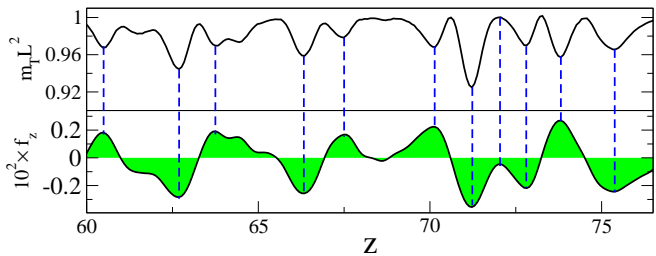}
\caption{(color) Magnetization of the system at $t=2\tau_Q$
($q=0$). The dashed lines facilitate observation of 
extrema coincidences. Results come from the same
simulation as in Figs. \ref{fig1}, \ref{fig2}; see \cite{units} for units. 
}
\label{fig3}
\end{figure}

Are different stages of this evolution 
experimentally observable? Let's look at $\tau_Q=10$ case
presented in Figs. \ref{fig1}-\ref{fig3}. The evolution from the phase
boundary to  
the first peak in  magnetization $M_T$ (the $q=0$ point) 
takes  $2.81\times37ms\cong104ms$ ($2\tau_Q=740ms$). 
Both these time scales are
well within the reach of the experiment  \cite{nature_kurn}.

\begin{figure}[t]
\includegraphics[width=\columnwidth,clip=true]{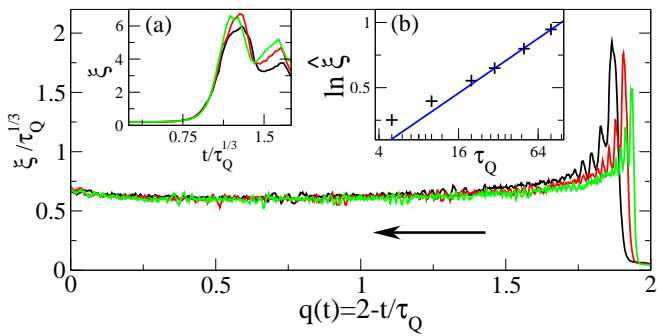}
\caption{(color) Dynamics of magnetic domains in $f_z$. Black line ($\tau_Q=30$),
         red line ($\tau_Q=50$), green line ($\tau_Q=80$). The arrow show direction 
	 of evolution on the main plot.
	 Inset (a): early stages of $\xi(t)$ evolution. 
	 Inset (b): dependence of the  typical post-transition 
	 domain size, $\hat\xi$,  on quench time. The fit was done to  $\tau_Q\ge30$ data 
	 (see text for details).
	 The figure shows  results averaged over $N_r=44$ runs; 
	 see \cite{units} for 	 units. 
}
\label{fig4}
\end{figure}

{\it The longitudinal magnetization.}
Initially,  $f_z(z)\approx0$ so that $\int dz f_z\approx0$. 
The conservation of the latter  allows only 
for creation of a network
of magnetic domains (nontopological structures with fixed $f_z$ sign)
having opposite polarizations. 
The domains 
appear by the time when the system enters unstable evolution and the
maxima of $|f_z|$ tend to move towards the minima of $m_T$
(Fig. \ref{fig3}). 
More quantitatively, 
we performed $N_r$ evolutions starting from different initial conditions, but
fixed $\sigma$. As in the experiment \cite{nature_kurn}, we average over these
runs to wash out shot-to-shot fluctuations.  
In Fig. \ref{fig4} we plot the mean domain size: 
$\xi=\sum_i\xi_z(i)/N_r$, where $i=1,...,N_r$ and
$\xi_z(i)$ is the mean domain size in the $i$-th run.
As shown in Fig. \ref{fig4}a, for $t\lesssim\hat t$ we observe 
$\xi\approx f(t/\tau_Q^{1/3})$  as for  $M_T(t)$.
The domains are formed on a time scale of $\sim\hat t$. 
A simple analysis based on  KZ theory \cite{kibble,zurek} suggests that
their characteristic {\it post-transition} size, $\hat \xi$, should be 
roughly given by $\int_0^{\sim\hat t} dt v_s(t)$, 
where $v_s(t)$ is a sound velocity. 
There are  two sound modes  in the broken-symmetry phase that propagate both spin and density 
fluctuations \cite{ueda_broken}: the faster (slower) one
has  velocity $\sim\sqrt{c_0}$ ($\sim\sqrt{q|c_1|}$). 
Putting any of these as $v_s$ into the integral, and 
assuming $\tau_Q\gg1$ for the slower mode, we get 
\begin{equation}
\label{XI}
\hat\xi\sim\tau_Q^{1/3}.
\end{equation}
This result correctly predicts the scaling 
property of the size of {\it post-transition} ``defects'' as is evident from 
the overlap of different curves in Fig. \ref{fig4}, which 
shows up for $\tau_Q\ge25$ or $0.9s$. Quantitatively,
we define  $\hat\xi$ as the value of $\xi$ averaged over $q\in[1/2,1]$ 
to wash out post-transition fluctuations. 
A fit got us $\ln\hat\xi= (-0.38\pm0.03)+ (0.30\pm0.01)\ln\tau_Q$,
in good agreement with (\ref{XI}). The fit was done to
$\tau_Q\ge30$ data and is presented in Fig. \ref{fig4}b which  illustrates 
that smaller $\tau_Q$ data  gradually departs from $1/3$ scaling law.

Now we  focus on the analytical calculations providing predictions
about  early stages of time-evolution. We assume 
that the wave-function stays close to the polar phase GS, 
$
\Psi^T=(\delta\psi_1(t), 1/\sqrt{L}+\delta\psi_0(t), \delta\psi_{-1}(t))
\exp(-i\mu t)
$,
where the chemical potential is $\mu=c_0/L$, $|\delta\psi_m|\ll1/\sqrt{L}$,
and $\int dz(\delta\Psi_0+\delta\Psi_0^*)\equiv0$ to keep $\int dz
\Psi^\dag\Psi=1+O(\delta\Psi^2)$.
Linearizing  
the coupled nonlinear-Schr\"odinger equations that describe the system
we get  $f_\chi={\rm Re} G_\chi$, where
$\chi=x,y$, $G_x= \sqrt{2}(\delta\Psi_1+\delta\Psi_{-1})/\sqrt{L}$, 
$G_y= i\sqrt{2}(\delta\Psi_1-\delta\Psi_{-1})/\sqrt{L}$, and 
$$
i\frac{d}{dt}G_\chi=
-\frac{1}{2}\frac{d^2}{dz^2}G_\chi+\frac{\alpha}{2}qG_\chi-\frac{\alpha}{2}(G_\chi+G_\chi^*),
$$
where $\alpha=2|c_1|/L$. 
To solve this equation we go to momentum space, $a_\chi(k)=\int dz
f_\chi\exp(ikz)$ and $b_\chi(k)=\int dz {\rm Im}G_\chi\exp(ikz)$,
getting 
\begin{equation}
\label{ewol}
\frac{d}{dt}
\left[\begin{array}{c}
a_\chi \\ b_\chi
\end{array}
\right]=\frac{1}{2}
\left(
\begin{array}{cc}
0 &  k^2+\alpha q \\
2\alpha-k^2-\alpha q & 0
\end{array}
\right)
\left[\begin{array}{c}
a_\chi \\ b_\chi
\end{array}
\right].
\end{equation}
Diagonalizing the matrix from Eq. (\ref{ewol}) we see that there is
instability  for $k^2/\alpha<2-q$  as in the 
Bogolubov spectrum of this model \cite{ueda_broken}.  
Thus, the system is stable in  the polar
phase, and so small initial perturbations do not grow during
the evolution towards broken-symmetry phase. 
The instability for $q<2$ is responsible for the magnetization jump 
 in Fig. \ref{fig1} and the subsequent breakdown of the linear approach (Fig.
 \ref{fig1}b).

To solve Eq. (\ref{ewol}) with $q(t)$ given by (\ref{q_od_t}) we derive the 
equation for $d^2a_\chi(t)/dt^2$, keep leading order terms 
in the slow transition ($\tau_Q\gg1$) 
and long-wavelength \protect{($k^2/\alpha\ll2$)} limits, 
and get  that 
\begin{equation}
\label{aki}
a_\chi(k,t)= \alpha_{k\chi}{\rm Ai}(s)+ \beta_{k\chi}{\rm Bi}(s), \ 
\frac{s}{\kappa}= \frac{t}{\tau_Q^{1/3}}-\frac{k^2\tau_Q^{2/3}}{\alpha},
\end{equation}
where $\kappa=(\alpha^2/2)^{1/3}$, 
$\alpha_{k\chi}$ and $\beta_{k\chi}$ are constants given by initial
conditions, while Ai and Bi are Airy functions.
From (\ref{aki}) we see that the instability  
arises from unbounded increase of the Bi$(s)$ function happening for 
$s>0$, i.e., $k^2/\alpha<2-q(t)$, which is a
dynamical manifestation of the static result for unstable modes.
This solution  works till $t\sim\hat t\sim \tau_Q^{1/3}$ when 
a significant increase of $f_\chi$ 
invalidates the linearized theory:  this calculation 
rigorously derives scaling  (\ref{t_hat}).  Additionally, the solution (\ref{aki}) 
can be reliably used as long as $\tau_Q\gg1$ or $37ms$, which is also supported 
by  numerics (Fig. \ref{fig1}a). The quench time scale in the experiment \cite{nature_kurn} 
is much smaller than this bound. Finally, these results hold for any initial state 
spread over the  $k$ modes.

The (re)scalings $t/\tau_Q^{1/3}$ and  
$\hat\xi\sim\hat t\sim\tau_Q^{1/3}$ derived above in a 1D system 
were also found by  different means in a 2D spinor condensate \cite{austen}.
A trivial extension of our mean-field analytical calculations to 2D and 3D systems
shows that they  hold for any number of spatial dimensions.

To summarize, we have developed a theory of the dynamics of symmetry-breaking 
in the quantum phase transition inspired by the experiment \cite{nature_kurn},
but for the range of quench rates that are sufficiently slow so that the
critical scalings can determine phase transition dynamics. This regime should
be accessible by a ``slower'' version of the quench \cite{nature_kurn}. 
Our analysis points to  a
Kibble-Zurek-like  scenario, where the state of the system departs from the
old symmetric vacuum with a delay $\sim\hat t$ after the critical point  was
crossed. This sets up an initial post-transition state with a characteristic 
length scale $\hat\xi$ (\ref{XI}). This scale should determine the initial density of 
topological features. In our 1D simulations  textures  appear, but we predict 
that in real 3D experiments other topological defects are created 
(as they were in \cite{nature_kurn}),
and the distance between them should be initially $\sim\hat\xi$. Such
topological defects are more stable than textures so measurement of their
density should be possible and would be a good test of the theory we have
presented.


\begin{thebibliography}{99}
\bibitem{units} 
The experiment \cite{nature_kurn} is done in the harmonic potential 
$m\omega_z^2(\lambda_x^2x^2+\lambda_y^2y^2+z^2)/2$, where $m$ is the
$^{87}$Rb mass, 
$\omega_z=2\pi\times4.3{\rm Hz}$, $\lambda_x=13$, and $\lambda_y=81.4$.
We use the   oscillator units through the paper for  
time ($1/\omega_z=37ms$), length ($\sqrt{\hbar/m\omega_z}=5.1\mu m$), and 
energy ($\hbar\omega_z$), and 
assume that the system stays in the harmonic oscillator ground states 
in $x$ and $y$ directions. Then, after skipping constant terms,
the 3D energy functional \cite{fer} reduces to dimensionless (\ref{E}) 
with the additional $\Psi^\dag\Psi z^2/2$ term, and 
$c_i=2N\alpha_i\sqrt{m\omega_z/\hbar}\sqrt{\lambda_x\lambda_y}/3$ 
($N=2\times10^6$ is the atom number, 
$\alpha_0= 16nm$, and  $\alpha_1=-\alpha_0/216.1$).
To approximate such a system by a box we assume that 
the total density of atoms in the harmonic trap center is the same as in the
box of size $L$. Neglecting in the Thomas-Fermi limit  
the first and the last term of (\ref{E}) we  get 
$L=2(\sqrt{c_0}2/3)^{2/3}\approx78$, i.e., $78\times5.1\mu m\approx0.4 mm$.
\end{thebibliography}
\end{document}